\documentclass[aps,prc,preprint,amsmath,amssymb,showpacs,preprintnumbers,superscriptaddress]{revtex4-1}
\usepackage{CJK}
\usepackage{graphicx}
\usepackage{dcolumn}
\usepackage{bm}
\usepackage{tabularx}
\usepackage{multirow}

\begin{document}

\title{Nuclear matrix elements of neutrinoless double-$\beta$ decay in the triaxial projected shell model}
\author{Y. K. Wang}
\affiliation{State Key Laboratory of Nuclear Physics and Technology, School of Physics, Peking University, Beijing 100871, China}
\author{P. W. Zhao}
\affiliation{State Key Laboratory of Nuclear Physics and Technology, School of Physics, Peking University, Beijing 100871, China}
\author{J. Meng}
\email{mengj@pku.edu.cn}
\affiliation{State Key Laboratory of Nuclear Physics and Technology, School of Physics, Peking University, Beijing 100871, China}
\affiliation{Yukawa Institute for Theoretical Physics, Kyoto University, Kyoto 606-8502, Japan}

\date{\today}
\begin{abstract}
  The nuclear matrix elements of neutrinoless double-$\beta$ decay for nuclei $^{76}$Ge, $^{82}$Se, $^{100}$Mo, $^{130}$Te, and $^{150}$Nd are studied within the triaxial projected shell model, which incorporates simultaneously the triaxial deformation and quasiparticle configuration mixing.
  The low-lying spectra and the $B(E2:0^+\rightarrow2^+)$ values are reproduced well.
  The effects of the quasiparticles configuration mixing, the triaxial deformation, and the closure approximation on the nuclear matrix elements are studied in detail.
  For nuclei $^{76}$Ge, $^{82}$Se, $^{100}$Mo, $^{130}$Te, and $^{150}$Nd, the nuclear matrix elements are respectively reduced by the quasiparticle configuration mixing by 6\%, 7\%, 2\%, 3\%, and 4\%, and enhanced by the odd-odd intermediate states by 7\%, 4\%, 11\%, 20\%, and 14\%.
  Varying the triaxial deformation $\gamma$ from $0^\circ$ to $60^\circ$ for the mother and daughter nuclei, the nuclear matrix elements change by 41\%, 17\%, 68\%, 14\%, and 511\% respectively for $^{76}$Ge, $^{82}$Se, $^{100}$Mo, $^{130}$Te, and $^{150}$Nd, which indicates the importance of treating the triaxial deformation consistently in calculating the nuclear matrix elements.
\end{abstract}
\maketitle
\date{today}

\section{Introduction}
The neutrinoless double-$\beta$ ($0\nu\beta\beta$) decay is a nuclear weak process in which an even-even nucleus decays to its even-even neighbor by emitting only two electrons.
It is a lepton-number-violating process and provides a sensitive probe to explore the Majorana nature of the neutrino~\cite{Schechter1982Phys.Rev.D252951--2954}.
The $0\nu\beta\beta$ decay is also regarded as an effective tool to determine the hierarchy of the neutrino mass spectrum~\cite{Avignone2008Rev.Mod.Phys.80481--516,Engel2017ReportsonProgressinPhysics80046301}.
Due to its great importance in revealing information associated with the fundamental physics, the detection of $0\nu\beta\beta$ decay has become the goal of several experimental projects worldwide~\cite{Agostini2018Physicalreviewletters120132503,Aalseth2018Physicalreviewletters120132502, Alduino2018Physicalreviewletters120132501,Albert2018Physicalreviewletters120072701, Arnold2016PhysicalReviewD93112008, Xue2017ChinesePhysicsC046002,Ni2019ChinesePhysicsC113001}.

In the light-neutrino exchange mechanism, the half-life $T^{0\nu}_{1/2}$ of $0\nu\beta\beta$ decay connects directly with the effective neutrino mass~\cite{Tomoda1991ReportsonProgressinPhysics5453--126},
\begin{equation}\label{eq:half-life-inves}
 [T^{0\nu}_{1/2}]^{-1} = G_{0\nu}g_A^4(0)\Big|\frac{\langle m_{\beta\beta}\rangle}{m_e}\Big|^2|M^{0\nu}|^2.
\end{equation}
Here, $m_e$ is the electron mass, $\langle m_{\beta\beta}\rangle$ is the effective neutrino mass,
$g_A(0)$ is the axial-vector coupling constant, and $G_{0\nu}$ is the kinematic phase-space factor~\cite{Kotila2012Phys.Rev.C85034316}.
Obviously, the accurate determination of the nuclear matrix element (NME) $M^{0\nu}$ is crucial for extracting the $\langle m_{\beta\beta}\rangle$ from the experimental half-life.

The NME $M^{0\nu}$ depends on the decay operator and the nuclear many-body wavefunctions.
The decay operator is derived from the second-order weak Hamiltonian constructed by the charged nucleonic and leptonic currents~\cite{Meng2017InternationalJournalofModernPhysicsE1740020}.
The effects from the higher order terms and the two-body currents on the decay operator have been studied extensively~\cite{Siimkovic1999Phys.Rev.C60055502,Menendez2011Phys.Rev.Lett.107062501,Wang2018Phys.Rev.C98031301,Belley2021Phys.Rev.Lett.042502}, and the quality of nonrelativistic reduction of the decay operator is also examined within a fully relativistic framework~\cite{Song2014Phys.Rev.C90054309,Meng2017InternationalJournalofModernPhysicsE1740020}.
The nuclear many-body wavefunctions are obtained from various nuclear models including the configuration interaction shell model (SM)~\cite{Caurier2008Phys.Rev.Lett.100052503,Menendez2009NuclearPhysicsA818139-151}, the quasiparticle random phase approximation (QRPA)~\cite{Siimkovic2013Phys.Rev.C87045501,Fang2015Phys.Rev.C92044301}, the interacting boson model (IBM)~\cite{Barea2015Phys.Rev.C034304}, the projected Hartree-Fock-Bogoliubov model (PHFB)~\cite{Chaturvedi2008Phys.Rev.C78054302,Rath2010Phys.Rev.C82064310}, and the generator coordinate method (GCM) based on the relativistic~\cite{Song2014Phys.Rev.C90054309,Yao2015Phys.Rev.C024316,Song2017Phys.Rev.C95024305} and nonrelativistic~\cite{Rodriguez2010Phys.Rev.Lett.105252503,Vaquero2013Phys.Rev.Lett.111142501} density functional theories (DFTs), etc.
These models are restricted either by the model space or the many-body correlations, which leads to the fact that the predicted NMEs differ by a factor $2$-$3$~\cite{Engel2017ReportsonProgressinPhysics80046301}.

The triaxial projected shell model (TPSM) carries out the configuration mixing based on a Nilsson mean field with the angular momentum projection technique~\cite{HARA1995InternationalJournalofModernPhysicsE04637-785}.
It has been successfully applied to study the nuclear rotational excitations, including the backbending phenomena~\cite{Hara1991ZeitschriftfurPhysikAHadronsandNuclei33915--21}, the superdeformed rotational bands~\cite{Sun1999Phys.Rev.Lett.83686--689}, the signature inversion~\cite{Gao2006PhysicsLettersB634195-199}, the $\gamma$ bands~\cite{Sheikh2008Phys.Rev.C77034313}, and the chiral~\cite{Chen2017Phys.Rev.C96051303,Chen2018PhysicsLettersB785211-216,Wang2019Phys.Rev.C99054303} and wobbling~\cite{Wang2020PhysicsLettersB135246} rotations, etc.

The simplified version of TPSM, namely the PHFB in axial deformation case, is performed to study the NMEs of $0\nu\beta\beta$ decay~\cite{Chaturvedi2008Phys.Rev.C78054302,Rath2010Phys.Rev.C82064310}.
With the help of Pfaffian algorithm~\cite{Bertsch2012Phys.Rev.Lett.108042505,Hu2014PhysicsLettersB734162-166} to evaluate the matrix elements of many-body operators, the newly developed TPSM in Refs.~\cite{Chen2017Phys.Rev.C96051303, Wang2019Phys.Rev.C99054303} include triaxial deformation and the configuration interaction induced by the quasiparticle excitations beyond the HFB vacuum, and treat even-even and the odd-odd nuclear systems simultaneously.
This provides us an opportunity to study the effects of the triaxial deformation and quasiparticle configuration mixing, and the effects beyond the closure approximation on the NMEs.

In this paper, the TPSM is applied to investigate the NMEs of $0\nu\beta\beta$ decay for nuclei $^{76}$Ge, $^{82}$Se, $^{100}$Mo, $^{130}$Te and $^{150}$Nd.
The influence of the triaxial deformation, the quasiparticles configuration mixing, and the commonly used closure approximation will be examined.

\section{Theoretical framework}
\subsection{Decay operator}
The charged-current weak Hamiltonian for the  $0\nu\beta\beta$ decay is~\cite{Walecka2012Muonphysics2113--217}
\begin{equation}
  \mathcal{H}_{\mathrm{weak}}(x)=\frac{G_F\cos\theta_C}{\sqrt{2}}j^\mu(x)\mathcal{J}^\dag_\mu(x)+\mathrm{H.c.},
\end{equation}
where $G_F$ and $\theta_C$ are the Fermi constant and Cabbibo angle, respectively.
The standard leptonic current $j^\mu$ is
 \begin{equation}
   j^\mu = \bar{e}(x)\gamma^\mu(1-\gamma_5)\nu_e(x),
  \end{equation}
and the hadronic current is expressed in terms of nucleon field $\psi$ as
 \begin{equation}
   \mathcal{J}^\dag_\mu = \bar{\psi}(x)\left[g_V(q^2)\gamma_\mu+ig_M(q^2)\frac{\sigma_{\mu\nu}}{2m_p}q^\nu
   -g_A(q^2)\gamma_\mu\gamma_5-g_P(q^2)q_\mu\gamma_5\right]\tau_-\psi(x).
  \end{equation}
Here, $m_p$ is the nucleon mass, $q^\mu$ is the transferred momentum from hadrons to leptons, the isospin lowing operator $\tau_- \equiv (\tau_1-i\tau_2)/2$, and $\sigma_{\mu\nu} \equiv i/2 [\gamma_\mu,\gamma_\nu]$.
The momentum dependent form factors $g_V(q^2), g_M(q^2), g_A(q^2)$, and $g_P(q^2)$ incorporate the correction of the finite nucleon size, and are reduced to the vector, weak-magnetism, axial-vector, and induced pseudoscalar coupling constants respectively in the zero momentum transfer limit.
The detailed formulas of the form factors can be found in Ref.~\cite{Siimkovic1999Phys.Rev.C60055502}.

Assuming that $0\nu\beta\beta$ decay is mediated by the light Majorana neutrinos, adopting the  long-wave approximation for the outgoing electrons, and neglecting the small energy transferred between nucleons, the NME can be derived with the help of $S$-matrix in the framework of second-order perturbative theory~\cite{Song2014Phys.Rev.C90054309},
\begin{equation}\label{eq:NME}
  M^{0\nu} = \langle\Psi_D|\hat{\mathcal{O}}^{0\nu}|\Psi_M\rangle.
\end{equation}
Here, $|\Psi_M\rangle$ and $|\Psi_D\rangle$ are respectively the nuclear wavefuntions of the mother and daughter nuclei,
and the decay operator $\hat{\mathcal{O}}^{0\nu}$ reads
\begin{equation}\label{eq:operator-R}
  \hat{\mathcal{O}}^{0\nu} = \frac{4\pi R}{g_A^2(0)}\int\int d^3x_1d^3x_2\int\frac{d^3q}{(2\pi)^3}
  \frac{e^{i\bm{q}\cdotp(\bm{x}_1-\bm{x}_2)}}{|\bm{q}|}\sum_m\frac{\mathcal{J}^\dag_\mu(\bm{x}_1)
  |m\rangle\langle m|\mathcal{J}^{\mu\dag}(\bm{x}_2)}{|\bm{q}|+E_m-(E_M + E_D)/2}.
\end{equation}
To make the NME dimensionless, $R = 1.2\times A^{1/3}$ fm is introduced~\cite{Engel2017ReportsonProgressinPhysics80046301}.
The wavefunctions and energies of intermediate odd-odd nuclear states are denoted respectively by $|m\rangle$ and $E_m$.

To simplify the calculations, the closure approximation is adopted in most cases, in which $E_m$ is replaced by an average one $\bar{E}$~\cite{Vergados2012ReportsonProgressinPhysics75106301,Engel2017ReportsonProgressinPhysics80046301}.
In such a way, using the relation $\sum_m|m\rangle\langle m| = 1$, the decay operator becomes
\begin{equation}\label{eq:operator-R1}
  \hat{\mathcal{O}}^{0\nu} = \frac{4\pi R}{g_A^2(0)}\int\int d^3x_1d^3x_2\int\frac{d^3q}{(2\pi)^3}
  \frac{e^{i\bm{q}\cdotp(\bm{x}_1-\bm{x}_2)}}{|\bm{q}|}\frac{\mathcal{J}^\dag_\mu(\bm{x}_1)
  \mathcal{J}^{\mu\dag}(\bm{x}_2)}{|\bm{q}|+E_d},
\end{equation}
where $E_d = \bar{E}-(E_M + E_D)/2$ represents the average excitation energies and is always approximated by a simple empirical formula $E_d = 1.12\times A^{1/2}$~\cite{Kotila2012Phys.Rev.C85034316}.

In the TPSM calculations, it is necessary to reduce the decay operator in Eq.~\eqref{eq:operator-R1} to the nonrelativistic form.
By neglecting the small energy transferred between nucleons in the nonrelativistic expansion, the hadronic current
$\mathcal{J}^{\mu\dag}$ is reduced as~\cite{Siimkovic1999Phys.Rev.C60055502}
\begin{equation}\label{eq:operator-NR}
  \mathcal{J}^{\mu\dag}(\bm{x})\rightarrow\sum_{n=1}^A\tau_-^n[g^{\mu0}\mathcal{J}^0(\bm{q}^2)+
  g^{\mu k}\mathcal{J}^k_n(\bm{q}^2)]\delta(\bm{x}-\bm{x}_n),
\end{equation}
where $\bm{x}_n$ represents the coordinate of the $n$-th nucleon, and $\mathcal{J}^0(\bm{q})$ and $\mathcal{J}_n(\bm{q})$ are respectively
\begin{equation}\label{eq:operator-NR1}
  \begin{split}
    &\mathcal{J}^0(\bm{q})=g_V(\bm{q}^2),\\
    &\mathcal{J}_n(\bm{q})=-g_Mi\frac{\bm{\sigma}_n\times\bm{q}}{2m_p}+g_A(\bm{q}^2)\bm{\sigma}
    -g_P(\bm{q}^2)\frac{\bm{q}\bm{\sigma}_n\cdotp\bm{q}}{2m_p}.
  \end{split}
 \end{equation}
Substituting Eqs.~\eqref{eq:operator-NR} and \eqref{eq:operator-NR1} into Eq.~\eqref{eq:operator-R1}, one can get the decay operator in the nonrelativistic form,
\begin{equation}
  \begin{split}
    \hat{\mathcal{O}}^{0\nu} &=\frac{4\pi R}{g_A^2(0)}\int\int d^3x_1d^3x_2\int\frac{d^3q}{(2\pi)^3}
    \frac{e^{i\bm{q}\cdotp(\bm{x}_1-\bm{x}_2)}}{|\bm{q}|}\frac{\mathcal{J}^\dag_\mu(\bm{x}_1)
    \mathcal{J}^{\mu\dag}(\bm{x}_2)}{|\bm{q}|+E_d}\\
    &=\frac{4\pi R}{g_A^2(0)}\int\int d^3x_1d^3x_2\int\frac{d^3q}{(2\pi)^3}
    \frac{e^{i\bm{q}\cdotp(\bm{x}_1-\bm{x}_2)}}{|\bm{q}|(|\bm{q}|+E_d)}\\
    &\times\sum_{nm}[-h_F(\bm{q}^2)+h_{GT}(\bm{q}^2)\bm{\sigma}_m\cdotp\bm{\sigma}_n-h_T(\bm{q}^2)S_{nm}]\tau^n_-\tau^m_-\delta(\bm{x}_1-\bm{x}_m)\delta(\bm{x}_2-\bm{x}_n),
  \end{split}
\end{equation}
with
\begin{equation}
  S_{mn} = 3(\bm{\sigma}_n\cdotp\hat{\bm{q}})(\bm{\sigma}_m\cdotp\hat{\bm{q}})-\bm{\sigma}_m\cdotp\bm{\sigma}_n.
\end{equation}
The terms $h_F(\bm{q}^2)$, $h_{GT}(\bm{q}^2)$, and $h_{T}(\bm{q}^2)$ respectively correspond to Fermi (F), Gamow-Teller (GT), and Tensor (T) momentum dependent couplings, i.e.~\cite{Siimkovic1999Phys.Rev.C60055502},
\begin{equation}
  \begin{split}
    h_F(\bm{q}^2) &=g_V^2(\bm{q}^2),\\
    h_{GT}(\bm{q}^2)&=g_A^2(\bm{q}^2)\left[1-\frac{2}{3}\frac{\bm{q}^2}{\bm{q}^2+m_\pi^2}+\frac{1}{3}
    \left(\frac{\bm{q}^2}{\bm{q}^2+m_\pi^2}\right)^2\right]+\frac{2}{3}\frac{g_M^2(\bm{q}^2)\bm{q}^2}{4m_p^2},\\
    h_T(\bm{q}^2)&=g_A^2(\bm{q}^2)\left[\frac{2}{3}\frac{\bm{q}^2}{\bm{q}^2+m_\pi^2}-\frac{1}{3}
    \left(\frac{\bm{q}^2}{\bm{q}^2+m_\pi^2}\right)^2\right]+\frac{1}{3}\frac{g_M^2(\bm{q}^2)\bm{q}^2}{4m_p^2},
  \end{split}
\end{equation}
where $m_\pi$ and $m_p$ are pion and proton masses respectively.

\subsection{Nuclear wavefunctions}
The nuclear wavefuntions of the mother and daughter nuclei $|\Psi_M\rangle$ and $|\Psi_D\rangle$ in Eq.~\eqref{eq:NME} are obtained from the TPSM.
The Hamiltonian in the TPSM is~\cite{Ring2004}
\begin{equation}\label{eq:PPQQ}
  \hat{H} = \hat{H}_0-\frac{\chi}{2}\sum_\mu\hat{Q}^\dag_\mu\hat{Q}_\mu-G_M\hat{P}^\dag\hat{P}-G_Q\sum_\mu\hat{P}^\dag_\mu\hat{P}_\mu,
\end{equation}
which includes a spherical single-particle Hamiltonian, a quadrupole-quadrupole interaction, as well as a monopole and a quadrupole pairing interaction.
The intrinsic vacuum state $|\Phi_0\rangle$ is obtained by the following variational equation,
\begin{equation}
  \delta\langle\Phi_0|\hat{H}-\lambda_p\hat{N}_p-\lambda_n\hat{N}_n|\Phi_0\rangle = 0.
\end{equation}
The Lagrange multipliers $\lambda_p$ and $\lambda_n$ are determined respectively by the proton number $Z$ and neutron number $N$.

Based on the obtained intrinsic vacuum $|\Phi_0\rangle$, the two quasiparticle states $|\Phi_\kappa\rangle$ for even-even and odd-odd nuclei can be constructed as
\begin{equation}
  \begin{split}
    &\mathrm{even} \mbox{-} \mathrm{even}~\mathrm{nuclei}:\quad |\Phi_\kappa\rangle\in \{\hat{\beta}^\dag_{\nu_i}\hat{\beta}^\dag_{\nu_j}|\Phi_0\rangle, \hat{\beta}^\dag_{\pi_i}\hat{\beta}^\dag_{\pi_j}|\Phi_0\rangle\},\\
    &\mathrm{odd} \mbox{-} \mathrm{odd}~\mathrm{nuclei}:\quad |\Phi_\kappa\rangle\in \{\hat{\beta}_{\nu_i}^\dag\hat{\beta}^\dag_{\pi_j}|\Phi_0\rangle\},
  \end{split}
\end{equation}
where $\hat{\beta}^\dag_\pi$ ($\hat{\beta}_\pi$) and $\hat{\beta}^\dag_\nu$ ($\hat{\beta}_\nu$) are respectively the quasiparticle creation (annihilation) operators for proton and neutron.
The rotational symmetry restoration for intrinsic states $|\Phi_\kappa\rangle$ is realized by the angular momentum projection, which leads to the projected basis,
\begin{equation}
  \{\hat{P}^I_{MK}|\Phi_\kappa\rangle\},
\end{equation}
where the three-dimensional angular momentum projection operator~\cite{Ring2004} is
\begin{equation}
  \hat{P}^I_{MK} = \frac{2I+1}{8\pi^2}\int d\Omega D^{I\ast}_{MK}(\Omega)\hat{R}(\Omega).
\end{equation}
The diagonalization of the Hamiltonian in the symmetry restored projected basis leads to the Hill-Wheeler equation,
\begin{equation}\label{eq:Hill-Wheeler}
  \sum_{\kappa'K'}\{\langle\Phi_\kappa|\hat{H}\hat{P}^I_{KK'}|\Phi_{\kappa'}\rangle-E^{I}
  \langle\Phi_\kappa|\hat{P}^{I}_{KK'}|\Phi_{\kappa'}\rangle\}F^{I}_{\kappa'K'} = 0,
\end{equation}
where $\langle\Phi_\kappa|\hat{P}^I_{KK'}|\Phi_{\kappa'}\rangle$ and $\langle\Phi_\kappa|\hat{H}\hat{P}^I_{KK'}|\Phi_{\kappa'}\rangle$ are  respectively the norm matrix element and the energy kernel.
They can be calculated with the efficient Pfaffian algorithm~\cite{Bertsch2012Phys.Rev.Lett.108042505,Hu2014PhysicsLettersB734162-166}.
By solving the Hill-Wheeler equation in Eq.~\eqref{eq:Hill-Wheeler}, one can get the eigenvalues $E^{I}$ and the corresponding projected wavefunctions,
\begin{equation}
  |\Psi^{I}\rangle = \sum_{K\kappa}F^{I}_{\kappa K}\hat{P}^I_{MK}|\Phi_\kappa\rangle.
\end{equation}
The obtained projected wavefunctions are then used to calculate the NME $M^{0\nu}$ in Eq.~\eqref{eq:NME}.

\subsection{Calculation of the NME}
With the obtained projected wavefunctions and the decay operator, the NME $M^{0\nu}$ can be expressed as
\begin{equation}\label{eq:NME1}
  \begin{split}
    M^{0\nu}_\alpha=&\langle\Psi^I_D|\hat{\mathcal{O}}^{0\nu}_\alpha|\Psi^I_M\rangle\\
    =&\sum_{KK'}\sum_{\kappa\kappa'}\frac{2I+1}{8\pi^2}\int d\Omega D^{I\ast}_{K'K}(\Omega)\langle\Phi^D_{\kappa'}|\hat{\mathcal{O}}^{0\nu}_\alpha\hat{R}(\Omega)|\Phi^M_\kappa\rangle F^I_{K\kappa}F^{I\ast}_{K'\kappa'},
  \end{split}
\end{equation}
where $\alpha$ denotes Fermi, Gamow-Teller, or Tensor, and $\hat{P}^{I\dag}_{M'K'}$ commutes with the decay operator $\hat{\mathcal{O}}^{0\nu}_\alpha$ and $\hat{P}^{I\dag}_{M'K'}\hat{P}^I_{MK} = \hat{P}^I_{K'K}\delta_{MM'}$.

The rotational matrix element $\langle\Phi^D_{\kappa'}|\hat{\mathcal{O}}^{0\nu}_\alpha\hat{R}(\Omega)|\Phi^M_\kappa\rangle$ in the second-quantized form is
\begin{equation}\label{eq:R-Matrix-Element}
  \begin{split}
    &\langle\Phi^D_{\kappa'}|\hat{\mathcal{O}}^{0\nu}_\alpha\hat{R}(\Omega)|\Phi^M_\kappa\rangle\\
    =&\sum_{\mu\nu\delta\gamma}\langle\mu\nu|\mathcal{O}_\alpha(1,2)|\delta\gamma\rangle
    \langle\Phi^D_0|\hat{\beta}^D_b\hat{\beta}^D_a(\hat{c}^{\dag}_\mu\hat{c}^{\dag}_\nu)
    (\hat{d}_\gamma\hat{d}_\delta)\hat{R}(\Omega)\hat{\beta}^{M\dag}_c
    \hat{\beta}^{M\dag}_d|\Phi^M_0\rangle\\
    =&\sum_{\mu\nu\delta\gamma}\langle\mu\nu|\mathcal{O}_\alpha(1,2)|\delta\gamma\rangle
    \langle\Phi^D_0|\hat{\beta}^D_b\hat{\beta}^D_a(\hat{c}^{\dag}_\mu\hat{c}^{\dag}_\nu)(\hat{d}_\gamma
    \hat{d}_\delta)\tilde{\hat{\beta}}^{M\dag}_c\tilde{\hat{\beta}}^{M\dag}_d|\tilde{\Phi}^M_0\rangle,
  \end{split}
\end{equation}
where $\hat{c}^{\dag}$ denotes the proton creation operator, $\hat{d}$ denotes the neutron annihilation operator, and
$|\Phi^M_0\rangle$ and $|\Phi^D_0\rangle$ are respectively the intrinsic vacuum states of the mother and daughter nuclei.
The indices $\mu, \nu, \delta$, and $\gamma$ run over the bases which are the eigenstates $|Nljm\rangle$ of the
spherical harmonic oscillator potential in the present paper.
The detailed calculation of the two-body matrix element $\langle\mu\nu|\mathcal{O}_\alpha(1,2)|\delta\gamma\rangle$ can be found in Ref.~\cite{Song2014Phys.Rev.C90054309}.

For the evaluation of the rotational overlap $\langle\Phi^D_0|\hat{\beta}^D_b\hat{\beta}^D_a(\hat{c}^{\dag}_\mu\hat{c}^{\dag}_\nu)
(\hat{d}_\gamma\hat{d}_\delta)\tilde{\hat{\beta}}^{M\dag}_c\tilde{\hat{\beta}}^{M\dag}_d|\tilde{\Phi}^M_0\rangle$, according to the strategy in Ref.~\cite{Hu2014PhysicsLettersB734162-166}, the following matrix elements of $S^{(\pm)}$ and $C^{(\pm)}$ are defined,
\begin{equation}
  S^{(+)}_{\mu k} =
  \begin{cases}
    -\frac{\langle\Phi^D_0|\hat{z}_k\hat{c}^{\dag}_\mu|\tilde{\Phi}^M_0\rangle}{\langle\Phi^D_0|\tilde{\Phi}^M_0\rangle}, &\quad \hat{z}_k\in \hat{\beta}^D_b,\hat{\beta}^D_a\\
    \frac{\langle\Phi^D_0|\hat{c}^{\dag}_\mu\hat{z}_k|\tilde{\Phi}^M_0\rangle}{\langle\Phi^D_0|\tilde{\Phi}^M_0\rangle}, &\quad \hat{z}_k\in \tilde{\hat{\beta}}^M_c,\tilde{\hat{\beta}}^M_d\\
  \end{cases},
\end{equation}
\begin{equation}
  S^{(-)}_{\mu k} =
  \begin{cases}
    -\frac{\langle\Phi^D_0|\hat{z}_k\hat{d}_\mu|\tilde{\Phi}^M_0\rangle}{\langle\Phi^D_0|\tilde{\Phi}^M_0\rangle}, &\quad \hat{z}_k\in \hat{\beta}^D_b,\hat{\beta}^D_a\\
    \frac{\langle\Phi^D_0|\hat{d}_\mu\hat{z}_k|\tilde{\Phi}^M_0\rangle}{\langle\Phi^D_0|\tilde{\Phi}^M_0\rangle}, &\quad \hat{z}_k\in \tilde{\hat{\beta}}^M_c,\tilde{\hat{\beta}}^M_d\\
  \end{cases},
\end{equation}
\begin{equation}
  \begin{split}
    &C^{(+)}_{\mu\nu} = \frac{\langle\Phi^D_0|\hat{c}^{\dag}_\mu\hat{c}^{\dag}_\nu|\tilde{\Phi}^M_0\rangle}
    {\langle\Phi^D_0|\tilde{\Phi}^M_0\rangle},\quad
    C^{(-)}_{\mu\nu} = \frac{\langle\Phi^D|\hat{d}_\mu\hat{d}_\nu|\tilde{\Phi}^M_0\rangle}
    {\langle\Phi^D_0|\tilde{\Phi}^M_0\rangle}.
  \end{split}
\end{equation}
The rotational overlap can then be expressed as
\begin{equation}\label{eq:R-Matrix-Element1}
  \begin{split}
    & \langle\Phi^D_0|\hat{\beta}^D_b\hat{\beta}^D_a(\hat{c}^{\dag}_\mu\hat{c}^{\dag}_\nu)(
    \hat{d}_\gamma\hat{d}_\delta)\tilde{\hat{\beta}}^{M\dag}_c\tilde{\hat{\beta}}^{M\dag}_d|\tilde{\Phi}^M_0\rangle\\
    =&~C^{(+)}_{\mu\nu}C^{(-)}_{\gamma\delta}\mathrm{Pf}(X)\langle\Phi^D_0|\tilde{\Phi}^M_0\rangle\\
    & + \sum_{ij}(-1)^{i+j}\alpha_{ij}C^{(+)}_{\mu\nu}S^{(-)}_{\delta i}S^{(-)}_{\gamma j}
    \mathrm{Pf}(X\{i,j\})\langle\Phi^D_0|\tilde{\Phi}^M_0\rangle\\
    & + \sum_{ij}(-1)^{i+j}\alpha_{ij}C^{(-)}_{\delta\gamma}S^{(+)}_{\mu i}S^{(+)}_{\nu j}
    \mathrm{Pf}(X\{i,j\})\langle\Phi^D_0|\tilde{\Phi}^M_0\rangle\\
    & + \sum_{ijkl}\alpha_{ijkl}S^{(+)}_{\mu i}S^{(+)}_{\nu j}S^{(-)}_{\delta k}S^{(-)}_{\gamma l}
    \langle\Phi^D_0|\tilde{\Phi}^M_0\rangle,
  \end{split}
\end{equation}
where $\alpha_{ij} =1$ for $i < j$, $\alpha_{ij} = -1$ for $i > j$, and $\alpha_{ijkl}=\alpha_{ij}\alpha_{ik}\alpha_{il}\alpha_{jk}\alpha_{jl}\alpha_{kl}$.
The skew-symmetric matrix $X$ has a dimension $4\times4$ and its matrix element in the lower triangle is~\cite{Hu2014PhysicsLettersB734162-166}
\begin{equation}\label{eq:Xmatrix}
  X_{ij} = \frac{\langle\Phi^D_0|\hat{z}_{i}\hat{z}_{j}|\tilde{\Phi}^M_0\rangle}{\langle\Phi^D_0|\tilde{\Phi}^M_0\rangle},\quad i < j.
\end{equation}
The $X(\{i,j\})$ in Eq.~\eqref{eq:R-Matrix-Element1} represents a sub-matrix of $X$ obtained by removing the rows and columns of $i,j$.
The matrix elements $S^{(\pm)}_{\mu k}, C^{(\pm)}_{\mu k}$, and $X_{ij}$ can be evaluated by the formulas in Ref.~\cite{Hu2014PhysicsLettersB734162-166}.

\section{Numerical details}
In the following, the NMEs for $0\nu\beta\beta$ decay candidates $^{76}$Ge, $^{82}$Se, $^{100}$Mo, $^{130}$Te, and $^{150}$Nd are calculated by the TPSM.
Three harmonic-oscillator major shells around the Fermi surface are taken in the calculation for both protons and neutrons.
The strengths of the monopole pairing interaction for protons and neutrons are same as those used in Ref.~\cite{Chaturvedi2008Phys.Rev.C78054302}, $G_M^p = 30/A$ MeV and $G_M^n = 20/A$ MeV.
Similar as Refs.~\cite{Sun1994Phys.Rev.Lett.723483--3486,Sun1996PhysicsReports264375-391}, the strength of the quadrupole pairing interaction is chosen as $G_Q = 0.2\times G_M$.
The strength of the quadrupole-quadrupole interaction $\chi$ is associated with the quadrupole deformation parameters $(\beta,\gamma)$ by the self-consistent relation~\cite{HARA1995InternationalJournalofModernPhysicsE04637-785}.
For nuclei $^{100}$Mo, $^{130}$Te, and $^{150}$Nd as well as their daughter nuclei, the quadrupole deformation parameters $(\beta,\gamma)$ are taken from Refs.~\cite{Singh2007TheEuropeanPhysicalJournalA33375--388,Chaturvedi2008Phys.Rev.C78054302}, in which the NMEs of $0\nu\beta\beta$ decay for these three nuclei have been studied by the PHFB.
For nuclei $^{76}$Ge and $^{82}$Se as well as their daughter nuclei $^{76}$Se and $^{82}$Kr, the $(\beta,\gamma)$ values are obtained self-consistently by the calculations of the relativistic DFT~\cite{Zhao2010Phys.Rev.C82054319,Zhao2015Phys.Rev.C92034319,Wang2017Phys.Rev.C96054324,Wang2018Phys.Rev.C97064321}.

\section{Results and discussion}
\begin{figure*}[h]
  \centering
  \includegraphics[width=0.9\textwidth]{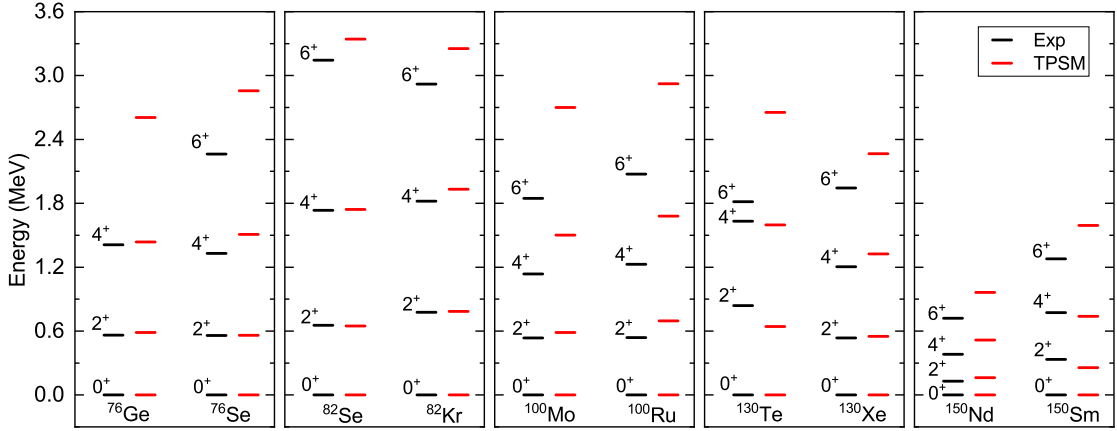}
  \caption{(Color online) Low-lying spectra for nuclei $^{76}$Ge, $^{76}$Se, $^{82}$Se, $^{82}$Kr, $^{100}$Mo, $^{100}$Ru, $^{130}$Te, $^{130}$Xe, $^{150}$Nd, and $^{150}$Sm calculated by the TPSM in comparison with the data~\cite{NNDC2.8}.}
  \label{Energy}
\end{figure*}

The low-lying spectra for nuclei $^{76}$Ge, $^{82}$Se, $^{100}$Mo, $^{130}$Te and $^{150}$Nd as well as their daughter nuclei $^{76}$Se, $^{82}$Kr, $^{100}$Ru, $^{130}$Xe and $^{150}$Sm calculated by the TPSM are shown in Fig.~\ref{Energy}, in comparison with the data.
It is found that the level schemes are reproduced satisfactorily by the TPSM calculations, especially for the $2^+_1$ states.
For higher spins with $I = 4, 6\hbar$, the calculated energy levels are slightly stretched compared with the data.
The reason for the deviation might be due to the neglect of the four quasiparticle configurations in the present calculation.
It is expected that the inclusion of those configurations would lower these states and work in this direction is in progress.

\begin{table*}[h]
  \centering
  \caption{$B(E2: 0^+_1\rightarrow 2^+_1)$ values (in $e^2$b$^2$) for nuclei $^{76}$Ge, $^{76}$Se, $^{100}$Mo, $^{82}$Se, $^{82}$Kr, $^{100}$Ru, $^{130}$Te, $^{130}$Xe, $^{150}$Nd, and $^{150}$Sm calculated by the TPSM in comparison with the data~\cite{Pritychenko2016AtomicDataandNuclearDataTables1071-139}.}
  \label{Table1}
  \begin{tabular}{p{1.2cm}|p{1.2cm}p{1.2cm}|p{1.2cm}p{1.2cm}|p{1.2cm}p{1.2cm}|p{1.2cm}p{1.2cm}|p{1.2cm}p{1.2cm}}
  \hline\hline
     &$^{76}$Ge& $^{76}$Se & $^{82}$Se& $^{82}$Kr& $^{100}$Mo & $^{100}$Ru & $^{130}$Te & $^{130}$Xe & $^{150}$Nd & $^{150}$Sm \\
  \hline
  TPSM &0.218& 0.304& 0.199& 0.210& 0.584 & 0.457 & 0.269 & 0.496 & 3.018& 2.321 \\
  Exp &0.278& 0.419& 0.180& 0.225& 0.530 & 0.493 & 0.296 & 0.634 & 2.707& 1.347 \\
  \hline\hline
  \end{tabular}
\end{table*}

The $E2$ transition probabilities $B(E2: 0_1^+\rightarrow2_1^+)$ calculated by the TPSM are shown in Table.~\ref{Table1}, in comparison with the data~\cite{Pritychenko2016AtomicDataandNuclearDataTables1071-139}.
The TPSM calculations reproduce the experimental $B(E2: 0_1^+\rightarrow2_1^+)$ values well except for $^{150}$Sm, which might be associated with a slightly large quadrupole deformation parameter $\beta$ adopted in the present TPSM calculation.

\begin{table}[h]
  \centering
  \caption{The NMEs of $0\nu\beta\beta$ decay calculated by the TPSM ($M^{0\nu}$) and TPHFB ($M^{0\nu'}$), together with their differences $\Delta M^{0\nu}$ for nuclei $^{76}$Ge, $^{82}$Se, $^{100}$Mo, $^{130}$Te, and $^{150}$Nd.
  The contributions from Gamow-Teller ($M^{0\nu}_{GT}, M^{0\nu'}_{GT}$), Fermi ($M^{0\nu}_{F}, M^{0\nu'}_{F}$) and Tensor ($M^{0\nu'}_{T}, M^{0\nu'}_{T}$) transitions, and the quadrupole deformation parameters ($\beta,\gamma$) adopted for the mother and daughter nuclei are also listed.}
  \begin{tabular}{cccccccccccccc}
  \hline\hline
  \multirow{2}{*}{Decay process} & \multicolumn{2}{c}{($\beta,\gamma$)} & & \multicolumn{4}{c}{TPSM} & & \multicolumn{4}{c}{TPHFB} &\multirow{2}{*}{$\Delta M^{0\nu}$}\\ \cline{2-3} \cline{5-8}\cline{10-13}
   &Mother & Daughter& &$M^{0\nu}$ & $M^{0\nu}_{GT}$ & $M^{0\nu}_F$ & $M^{0\nu}_T$ & &$M^{0\nu'}$& $M^{0\nu'}_{GT}$& $M^{0\nu'}_F$ & $M^{0\nu'}_{T}$ &  \\
  \hline
     $^{76}$Ge  $\rightarrow$ $^{76}$Se  &($0.18,0^\circ$) &($0.22,60^\circ$)& &3.17&2.67 & -0.72 & -0.01& &3.37 & 2.84 & -0.77 & -0.01 &0.20\\
     $^{82}$Se  $\rightarrow$ $^{82}$Kr  &($0.17,0^\circ$) &($0.14,0^\circ$) & &2.59&2.16 & -0.59 & -0.02& &2.78 & 2.32 & -0.63 & -0.02 &0.19\\
     $^{100}$Mo $\rightarrow$ $^{100}$Ru &($0.23,0^\circ$) &($0.21,0^\circ$) & &3.92&3.46 & -0.78 & -0.03& &3.99 & 3.52 & -0.79 & -0.03 &0.07\\
     $^{130}$Te $\rightarrow$ $^{130}$Xe &($0.12,0^\circ$) &($0.17,0^\circ$) & &2.92&2.64 & -0.56 & -0.01& &3.00 & 2.71 & -0.58 & -0.01 &0.08\\
     $^{150}$Nd $\rightarrow$ $^{150}$Sm &($0.28,0^\circ$) &($0.24,0^\circ$) & &3.29&2.89 & -0.55 & -0.02& &3.44 & 3.02 & -0.58 & -0.02 &0.15\\
     \hline\hline
   \end{tabular}\label{Tab:NMEs}
\end{table}

The satisfactory reproduction of the low-lying spectra and $B(E2: 0_1^+\rightarrow2_1^+)$ values gives us confidence to apply TPSM to study the $0\nu\beta\beta$ decay.
The NME $M^{0\nu}$ and the contributions from Gamow-Teller ($M^{0\nu}_{GT}$), Fermi $(M^{0\nu}_F)$ and Tensor ($M^{0\nu}_T$) transitions calculated by the TPSM for nuclei $^{76}$Ge, $^{82}$Se, $^{100}$Mo, $^{130}$Te, and $^{150}$Nd are shown in the 4th to 7th columns of Table.~\ref{Tab:NMEs}.
The main contribution of the NME comes from the Gamow-Teller $M^{0\nu}_{GT}$, which exhausts 85\% of the total NME.
The contributions of Fermi and Tensor transitions to the total NME are around $14\%$ and $1\%$, respectively. Therefore, ignoring the Tensor contribution in the TPSM calculations can be a good approximation.

In order to study the effects of the quasiparticle configuration mixing, the NME $M^{0\nu'}$ calculated by the triaxial PHFB (TPHFB) are shown in the 8th and 11th columns in Table.~\ref{Tab:NMEs}.
The last column shows the differences $\Delta M^{0\nu}$ between $M^{0\nu'}$ and $M^{0\nu}$, which reveals the effect of quasiparticle configuration mixing missing in the TPHFB.
It is found that the quasiparticle configuration mixing reduces the NMEs ranging from 2\% to 7\%.

\begin{figure*}[h!]
  \centering
  \includegraphics[width=0.5\textwidth]{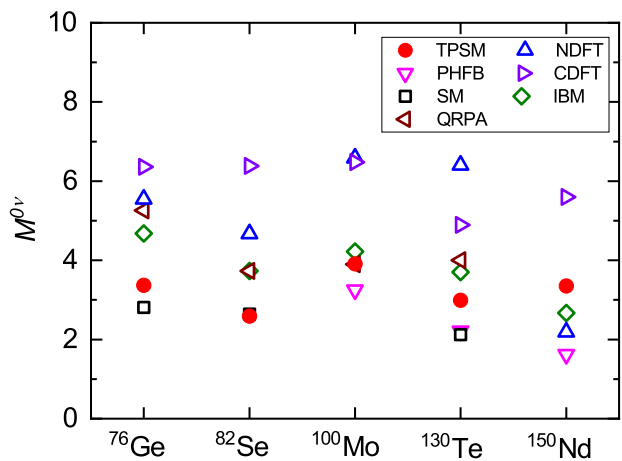}
  \caption{(Color online) The NMEs of $0\nu\beta\beta$ decay calculated by the TPSM in comparison with those from the nonrelativistic DFT (NDFT)~\cite{Vaquero2013Phys.Rev.Lett.111142501}, relativistic DFT (CDFT)~\cite{Song2017Phys.Rev.C95024305}, IBM~\cite{Barea2015Phys.Rev.C034304}, PHFB~\cite{Chaturvedi2008Phys.Rev.C78054302}, QRPA~\cite{Hyvaerinen2015Phys.Rev.C91024613}, and SM~\cite{Horoi2016Phys.Rev.C93024308}.}
  \label{Fig:NMEs-compare}
\end{figure*}

The NME $M^{0\nu}$ for nuclei $^{76}$Ge, $^{82}$Se, $^{100}$Mo, $^{130}$Te, and $^{150}$Nd calculated by the TPSM are compared with the ones from the NDFT~\cite{Vaquero2013Phys.Rev.Lett.111142501}, CDFT~\cite{Song2017Phys.Rev.C95024305}, IBM~\cite{Barea2015Phys.Rev.C034304}, PHFB~\cite{Chaturvedi2008Phys.Rev.C78054302}, QRPA~\cite{Hyvaerinen2015Phys.Rev.C91024613}, and SM~\cite{Horoi2016Phys.Rev.C93024308}, as shown in Fig.~\ref{Fig:NMEs-compare}.
The  $M^{0\nu}$ in TPSM calculations are larger than those from the PHFB and SM, and smaller than the results given by the NDFT and CDFT calculations.
Compared with the PHFB calculation in Ref.~\cite{Chaturvedi2008Phys.Rev.C78054302}, the valence space of the TPSM is chosen as three major shells, and the contribution from quasiparticle configuration mixing beyond the HFB vacuum has been taken into account.

\begin{figure*}[h]
  \centering
  \includegraphics[width=0.75\textwidth]{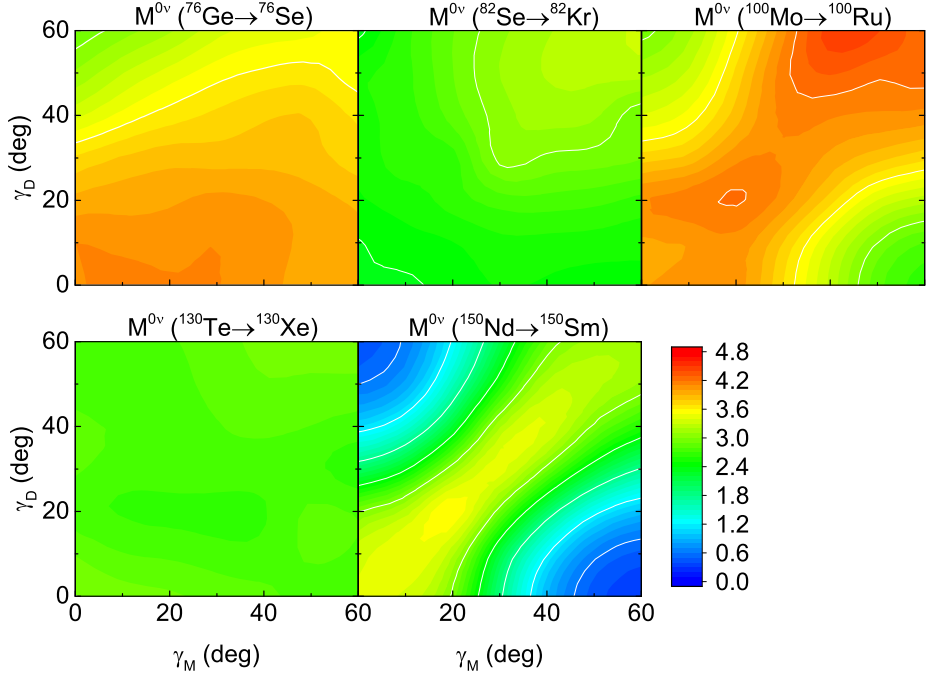}
  \caption{(Color online) The NMEs of $0\nu\beta\beta$ decay for nuclei $^{76}$Ge, $^{82}$Se, $^{100}$Mo, $^{130}$Te, and $^{150}$Nd as functions of the triaxial deformation parameters for the mother and daughter nuclei.}
  \label{Fig:NMEs-triaxial}
\end{figure*}

\begin{figure*}[h]
  \centering
  \includegraphics[width=0.75\textwidth]{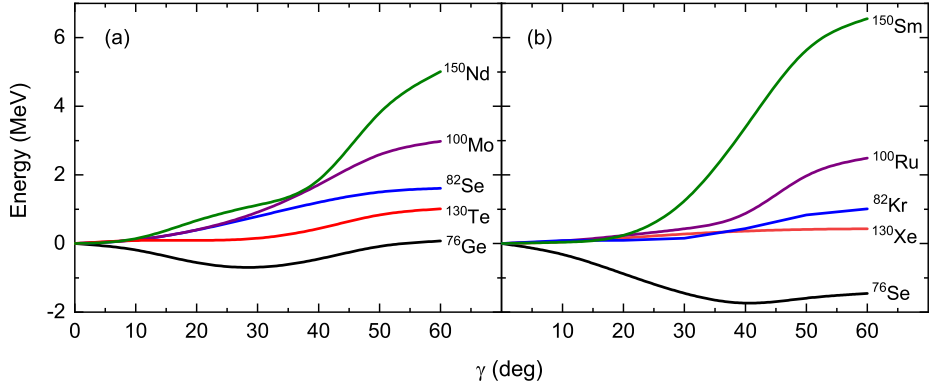}
  \caption{(Color online) Potential energy curves of $0^+$ states as functions of the triaxial deformation parameters $\gamma$ for nuclei $^{76}$Ge, $^{82}$Se, $^{100}$Mo, $^{130}$Te, and $^{150}$Nd (a) as well as their corresponding daughter nuclei (b) in the TPSM.}
  \label{Fig:Energy-gamma}
\end{figure*}

The NMEs shown in Table.~\ref{Tab:NMEs} are obtained by assuming that the nuclei under consideration are all axially deformed.
It is interesting to explore the evolution of NMEs with the nuclear triaxial deformation.
In Fig.~\ref{Fig:NMEs-triaxial}, the NMEs of $0\nu\beta\beta$ decay for nuclei $^{76}$Ge, $^{82}$Se, $^{100}$Mo, $^{130}$Te, and $^{150}$Nd as functions of the triaxial deformation parameters for the mother ($\gamma_M$) and daughter ($\gamma_D$) nuclei are shown.
For nuclei $^{82}$Se and $^{130}$Te, the NMEs remain roughly unchanged with the triaxial deformation parameters.
This can be explained by the potential energy curves of $0^+$ states shown in Fig.~\ref{Fig:Energy-gamma}.
The potential energy curves of $0^+$ states for $^{82}$Se and $^{130}$Te, together with their corresponding daughter nuclei $^{82}$Kr and $^{130}$Xe, are rather soft, which indicates the corresponding projected wavefunctions are not sensitive to the triaxial deformation.
This explains the nearly unchanged NMEs for $^{82}$Se and $^{130}$Te with $\gamma_M$ and $\gamma_D$.

For nuclei $^{100}$Mo and $^{150}$Nd, the NMEs depend sensitively on the triaxial deformation parameters.
The mother and daughter nuclei with similar triaxial deformation parameters trend to give larger $M^{0\nu}$.
The variation of $M^{0\nu}$ with $\gamma_M$ and $\gamma_D$ can be explained by the stiffness of the potential energy curves for $^{100}$Mo and $^{150}$Nd, together with their corresponding daughter nuclei $^{100}$Ru and $^{150}$Sm, as shown in Fig.~\ref{Fig:Energy-gamma}.

Although $^{76}$Ge and $^{76}$Se are axially deformed in the relativistic DFT calculations, the rotational symmetry restored states $0^+$ in the TPSM for $^{76}$Ge and $^{76}$Se are triaxially deformed with $\gamma_M = 30^\circ$ and $\gamma_D = 40^\circ$, as shown in Fig.~\ref{Fig:Energy-gamma}.
With these corresponding triaxial deformation parameters, the resulting $M^{0\nu}$ is $3.72$ which is $17\%$ larger than the value $3.17$ in the axial deformation case shown in Table.~\ref{Tab:NMEs}.
This indicates the importance of treating the triaxial deformation correctly when calculating the NMEs of $0\nu\beta\beta$ decay.

\begin{figure*}[h]
  \centering
  \includegraphics[width=0.8\textwidth]{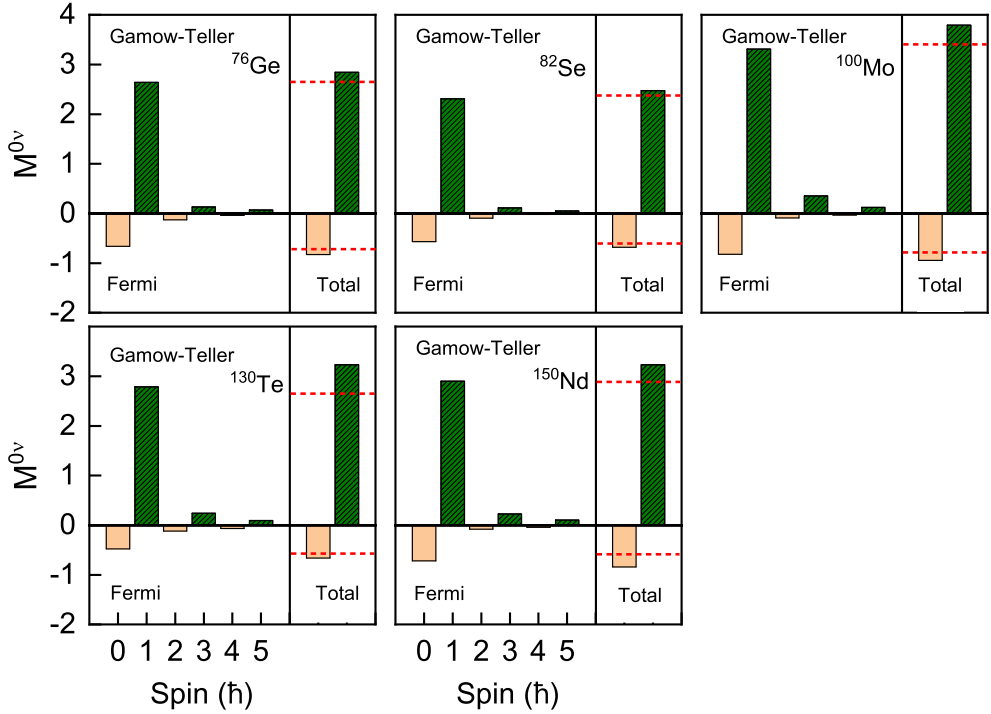}
  \caption{(Color online) The NMEs of $0\nu\beta\beta$ decay for nuclei $^{76}$Ge, $^{82}$Se, $^{100}$Mo, $^{130}$Te, and $^{150}$Nd calculated by the TPSM with odd-odd intermediate states at different spin.
  The left panel in each sub-figure denotes the contribution of the Gamow-Teller and Fermi transitions.
  The right panels denote the total Gamow-Teller and Fermi NMEs in comparison with the corresponding ones calculated by the closure approximation (dashed line).}
  \label{Fig:NMEs-nonclosure}
\end{figure*}

To simplify the calculations, the closure approximation is adopted in most of the previous calculations.
The present TPSM treats the even-even and odd-odd nuclei in a unified way.
We can investigate the effects of the closure approximation and study the contributions of odd-odd intermediate states to the NMEs.
In Fig.~\ref{Fig:NMEs-nonclosure}, the NME $M^{0\nu}$ calculated by the TPSM with contributions of odd-odd intermediate states at different spin for nuclei $^{76}$Ge, $^{82}$Se, $^{100}$Mo, $^{130}$Te, and $^{150}$Nd are shown, in comparison with the results from the closure approximation.
The Gamow-Teller and the Fermi NMEs for the intermediate states at different spin are denoted by bars, and the dashed lines represent the results calculated by the closure approximation.
The Gamow-Teller NMEs are mainly contributed from the odd-odd intermediate states with $I = 1\hbar$, which exhausts more than $80\%$ of the total Gamow-Teller NMEs.
This can be understood by the Gamow-Teller operator $\sigma\tau_-$ and the initial and finial states $0^+$ for the mother and daughter nuclei.
The Fermi NMEs are mainly contributed from the intermediate states with $I = 0\hbar$.
The states with $I = 2, 4\hbar$ contribute less than $10\%$ to the total Fermi NMEs.

Comparing with the results in Table.~\ref{Tab:NMEs} with the closure approximation, the odd-odd intermediate states enhance the NMEs.
For nuclei $^{76}$Ge, $^{82}$Se, $^{100}$Mo, $^{130}$Te, and $^{150}$Nd, the contribution of the odd-odd intermediate states increase respectively the values of the NMEs by $7\%$, $4\%$, $11\%$, $20\%$, and $14\%$.
In comparison, an enhancement about 10\% is given by the SM calculations for $^{48}$Ca~\cite{Senkov2013Phys.Rev.C88064312}.

\section{Summary}
In summary, the nuclear matrix elements of neutrinoless double-$\beta$ decay for nuclei $^{76}$Ge, $^{82}$Se, $^{100}$Mo, $^{130}$Te, and $^{150}$Nd are studied within the triaxial projected shell model, which incorporates simultaneously the triaxial deformation and quasiparticle configuration mixing.
The low-lying spectra and the $B(E2:0^+\rightarrow2^+)$ values for nuclei under consideration are reproduced well.
The effects of the quasiparticles configuration mixing, the triaxial deformation, and the commonly used closure approximation are examined.

The inclusion of the quasiparticle configuration mixing in the configuration space reduces the nuclear matrix element ranging from 2\% to 7\%.
In comparison with the results by the closure approximation, the odd-odd intermediate states systematically enhance the nuclear matrix elements for nuclei $^{76}$Ge, $^{82}$Se, $^{100}$Mo, $^{130}$Te, and $^{150}$Nd by $7\%$, $4\%$, $11\%$, $20\%$, and $14\%$, respectively.

The mother and daughter nuclei with similar triaxial shape tend to give larger nuclear matrix elements.
After examining the nuclear matrix elements as functions of the triaxial deformation parameters, it is found the nuclear matrix elements for $^{76}$Ge, $^{82}$Se, $^{100}$Mo, $^{130}$Te, and $^{150}$Nd vary with $\gamma$ from $0^\circ$ to $60^\circ$ by 41\%, 17\%, 68\%, 14\%, and 511\% respectively.
This indicates the importance of treating the triaxial deformation consistently in calculating the nuclear matrix elements of neutrinoless double-$\beta$ decay.

Although $^{76}$Ge and $^{76}$Se are axially deformed in the calculations of the relativistic density functional theory, the rotational symmetry restored states $0^+$ in the triaxial projected shell model for $^{76}$Ge and $^{76}$Se are triaxially deformed.
With these corresponding triaxial deformation $\gamma = 30^\circ$ and $40^\circ$, the resulting nuclear matrix element is $3.72$, which is $17\%$ larger than the value $3.17$ in the axial deformation case.
Future work in developing the generator coordinate method is necessary to calculate the nuclear matrix elements for $^{76}$Ge by mixing the nuclear shapes.

\begin{acknowledgments}
This work was partly supported by the National Key R\&D Program of China (Contract No. 2018YFA0404400 and No. 2017YFE0116700), the National Natural Science Foundation of China (Grants No. 11935003, No. 11975031, No. 11875075, and No. 12070131001), the China Postdoctoral Science Foundation (2020M680183), and the High-performance Computing Platform of Peking University.
\end{acknowledgments}

%

\end{document}